\documentclass[pre,aps,12pt]{revtex4}
\usepackage{graphicx}

\begin{document}

\title{Spin triangle chain with ferromagnetic and antiferromagnetic interactions on
the transition line}
\author{V.~Ya.~Krivnov}
\email{krivnov@deom.chph.ras.ru}
\author{D.~V.~Dmitriev}
\affiliation{Institute of Biochemical Physics of RAS, Kosygin str.
4, 119334, Moscow, Russia.}
\date{}

\begin{abstract}
We investigate spin-$\frac{1}{2}$ anisotropic model of a linear
chain of triangles with competing ferro- and antiferromagnetic
interactions and ferromagneic Heisenberg interactions between
triangles. For a certain ratio between interactions the one-magnon
excitation band is dispersionless leading to an existence of
localized-magnon states which form macroscopically degenerated
ground states. The spectrum of excitations has a specific
structure depending on the value of the triangle-triangle
interaction. Such specific structure determines the
low-temperature thermodynamics and, in particular, the temperature
dependence of a specific heat. In the limit of strong anisotropy
of interactions the spectrum has a multi-scale structure which
consists of subsets rank-ordered on small parameter. Each subset
is responsible for the appearance of the peak in the temperature
dependence of the specific heat.
\end{abstract}

\maketitle

\section{Introduction}

Low-dimensional quantum magnets on geometrically frustrated
lattice have been extensively studied \cite{Diep, Lacrose}. An
important class of these systems includes lattice consisting of
triangles. For special relations between exchange interactions
such systems have a dispersionless one-magnon band (flat-band).
There is a wide class of the frustrated quantum antiferromagnets
(AF) in which a flat-band exists \cite{ModernPhysics, Shulen, Mac,
Zhit, Derzhko2004, Zhit2}. The one-magnon flat-band leads to an
existence of localized multi-magnon states which form a
macroscopically degenerate ground state in the saturation magnetic
field. Such models have specific low-temperature properties such
as an jump in the magnetization curve, an extra low-temperature
peak in the specific heat, nonzero residual entropy, etc.
\cite{Schmidt, Honecker, Zhitomir, Derzhko2004, Zhit2, Capponi,
Balika}.

\begin{figure}[tbp]
\includegraphics[width=4in,angle=0]{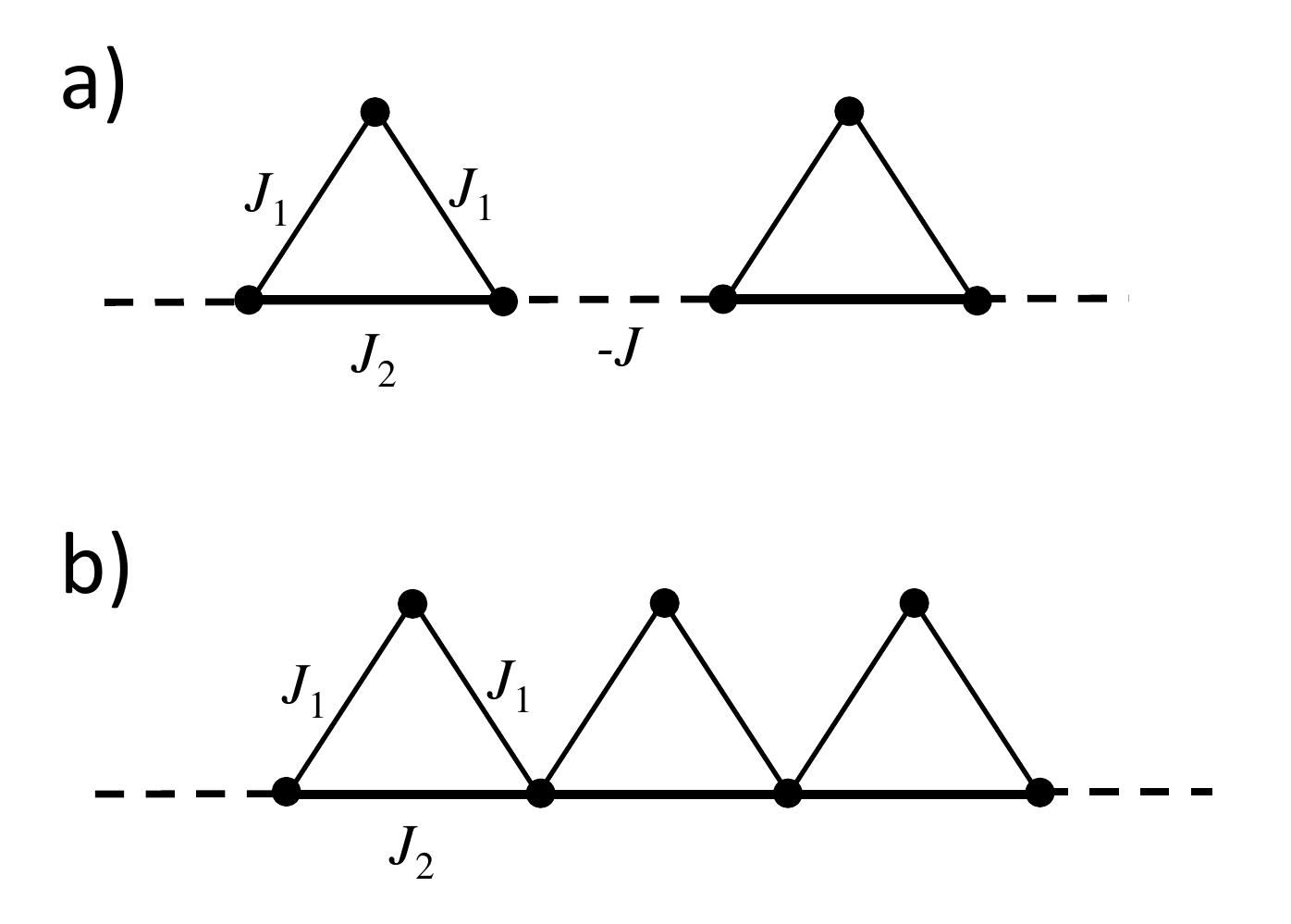}
\caption{(a) Model of linear chain of connected triangles and (b)
$\triangle$-chain model.} \label{triangles}
\end{figure}

An interesting one-dimensional example of such systems is
spin-$\frac{1}{2}$ delta-chain consisting of a linear chain of
connected triangles as shown in Fig.\ref{triangles}. The ground
state and the low-temperature properties of the AF delta-chain
with flat-band have been studied in detail \cite{Derzhko2004,
Derzhko2006, Zhitomir, ModPhys, Shulen, Zhit, Brenig} and it was
shown that it exhibits a number of peculiar properties.

Another class of the frustrated quantum systems with exact
localized multi-magnon states are the models with competing ferro-
and antiferromagnetic interactions (F-AF models). At definite
value of the ratio of these interactions the model has massively
degenerate ground state at zero magnetic field. The main
difference between AF and F-AF models with flat-band consists of
additional specifically overlapping localized multi-magnon
complexes which are exact ground states together with isolated
localized magnons. It leads to the macroscopical degeneracy of the
ground state and the residual entropy is larger than that for the
AF models. The example of this model is the same delta-chain. The
properties of the F-AF delta-chain has been studied in \cite{DK,
KD, ferri, DKRS, Schnack}. In these works the ground state
degeneracy has been determined and the relation of the low-energy
excitations to the low-temperature properties has been studied.

In this paper we study another example of the frustrated
spin-$\frac{1}{2}$ model with ferromagnetic and antiferromagnetic
interactions. This model represents a linear chain of triangles
connected by the ferromagnetic Heisenberg interactions
(spin-$\frac{1}{2}$ triangle chain model) as shown in
Fig.\ref{triangles}. The interaction $J_{1}$ between the basal and
the apical spins of triangle is ferromagnetic and the basal-basal
interaction $J_{2}$ is antiferromagnetic. The interaction $J$
between triangles is ferromagnetic. The Hamiltonian of this model
has a form
\begin{equation}
H=\sum_{i=1}^{n}H_{i}+\sum_{i=1}^{n}V_{i,i+1}  \label{H}
\end{equation}%
where%
\begin{eqnarray}
H_{i} &=&J_{1}\sum_{m=1,2}\left(
s_{i,m}^{x}s_{i,m+1}^{x}+s_{i,m}^{y}s_{i,m+1}^{y}+\Delta
_{1}(s_{i,m}^{z}s_{i,m+1}^{z}-\frac{1}{4})\right)   \label{H1} \\
&&+J_{2}\left( s_{i,1}^{x}s_{i,3}^{x}+s_{i,1}^{y}s_{i,3}^{y}+\Delta
_{2}(s_{i,1}^{z}s_{i,3}^{z}-\frac{1}{4})\right)   \nonumber \\
V_{i,i+1} &=&-J(\mathbf{s}_{i,3}\mathbf{s}_{i+1,1}-\frac{1}{4})  \nonumber
\end{eqnarray}%
where $n$ is a number of triangles, $\Delta _{1}$ and $\Delta
_{2}$ are parameters of the anisotropy of the basal-apical and the
basal-basal interactions ($\Delta _{1},\Delta _{2}>0$). The
constants in Eq.(\ref{H1}) are chosen so that the energy of the
ferromagnetic state with $S^{z}=\pm \frac{3}{2}n$ is zero. The
periodic boundary conditions are imposed. Further we put
$J_{1}=-1$ and $J_{2}=\alpha $.

Model (\ref{H}) is an extension of the F-AF delta chain. We will
show that the spin-$\frac{1}{2}$ triangle model (\ref{H}) has a
huge manifold of ground states and we focus on the studies of the
influence of the triangle-triangle ferromagnetic interactions on
the ground state properties and on the low-temperature
thermodynamics.

The paper is organized as follows. In Sec.II we derive the
necessary conditions on the Hamiltonian parameters for which the
model has exact localized magnon states. In Sec.III we study the
spin triangle model on the transition line between different
phases and represent the results for the ground state degeneracy.
Here we also study the structure of the spectrum and the density
of states and their relation to the low-temperature
thermodynamics. In Sec.IV we study the properties of the model
with strong anisotropy of interactions. In Sec.V we give a summary
of our results.

\section{Flat-band and localized magnons}

At first, we calculate the energy of one-magnon states, i.e. the
states in the spin sector $S^{z}=S_{\max }-1$ ($S_{\max
}=\frac{3}{2}n$). The one-magnon energy $E(k)$ can be found from
the equation
\begin{equation}
(E(k)-\Delta _{1})[(E(k)+A)^{2}-(E(k)+A)(\alpha +J)+\frac{\alpha J}{2}%
(1+\cos k)]-\frac{E(k)+A}{2}+\frac{J}{4}(1+\cos k)=0
\end{equation}%
where $A=\frac{1}{2}(\Delta _{1}-\alpha (1+\Delta _{2}))$.

There are three branches of the one-magnon energy and the lowest
branch $E_{1}(k)$ becomes dispersionless (flat band) under the
condition
\begin{equation}
\Delta _{2}=\frac{1}{\alpha ^{2}}-\frac{\Delta _{1}}{\alpha }-1
\label{dispers}
\end{equation}%
and $E_{1}(k)$ is%
\begin{equation}
E_{1}(k)=\Delta _{1}-\frac{1}{2\alpha }  \label{magnon energy}
\end{equation}

We notice that the condition (\ref{dispers}) and the energy
$E_{1}$ do not depend on $J$.

The energy of two other branches are%
\begin{equation}
E_{2,3}(k)=E_{1}+\frac{1}{2}\left[ \alpha +J+\frac{1}{2\alpha }\pm \sqrt{%
(\alpha -J+\frac{1}{2\alpha })^{2}+2\alpha J(1-\cos k)}\right]
\label{k=1 excitation}
\end{equation}%
and $E_{2,3}(k)>E_{1}$.

The one-magnon flat band means the existence of the localized
states which can be chosen as
\begin{equation}
\varphi _{m,m+1}\left\vert F\right\rangle =(s_{m,2}^{-}+\alpha
s_{m,3}^{-}+\alpha s_{m+1,1}^{-}+s_{m+1,2}^{-})\left\vert F\right\rangle
\qquad m=1,2,\ldots n  \label{flat}
\end{equation}%
where $\left\vert F\right\rangle $ is the ferromagnetic state with
all spins up and $s_{i}^{-}$ is the spin lowering operator.

The wave function (\ref{flat}) is localized in the valley between
neighboring triangles. Because one magnon is localized between $m$
and ($m+1$) triangles it is possible to construct $k\leq n/2$
non-overlapping localized (independent) magnons with the total
energy $E_{k}=kE_{1}$ and all of them are exact states. If
$E_{1}<0$ ($\Delta _{1}<\frac{1}{2\alpha }$) the lowest energy of
these states is realized in the spin sector $S^{z}=S_{\max }-n/2$.
The numerical calculations show that at $\Delta
_{1}<\frac{1}{2\alpha }$ the ground state of model (\ref{H}) is in
the spin sector $S^{z}=0 $. For $E_{1}>0$ the ground state is
ferromagnetic with zero energy.

\section{Spin triangle chain on the transition line}

The ferromagnetic state for model (\ref{H}) becomes unstable when
one-magnon excitations becomes negative. Therefore, the transition
line between the ferromagnetic and other (ferrimagnetic or
$S^{z}=0$) ground state phases is defined by the condition
$E_{1}=0$. In this case the model is described by one parameter
that is convenient to choose as $\Delta_1$, and according to
conditions (\ref{dispers}), (\ref{magnon energy}) two other
parameters are $\alpha =\frac{1}{2\Delta_1}$ and
$\Delta_2=2\Delta_1^2-1$. The model on the transition line has the
macroscopically degenerated ground state. To explain this fact let
us consider one separate triangle $H_{i}$. From eight eigenstates
of $H_{i}$ six states has zero energy: two states with $S^{z}=\pm
3/2$ and four states with $S^{z}=\pm 1/2$. The energy of two other
states with $S^{z}=\pm 1/2$ is positive and it is $E=\Delta
_{1}+\frac{1}{2\Delta _{1}}$. Because the Hamiltonians $H_{i}$ and
$V_{i,i+1}$ do not commute with each other the ground state energy
$E_{0}$ of $H$ satisfies an inequality
\begin{equation}
E_{0}\geq E_{0}(i)+E_{0}(i,i+1)=0  \label{inequality}
\end{equation}%
where $E_{0}(i)$ and $E_{0}(i,i+1)$ are the ground state energies
of $H_{i}$ and $V_{i,i+1}$ which are zero. Since independent
magnon states have zero energy, the inequality (\ref{inequality})
becomes the equality and all such states are ground states. In
Refs.\cite{DK, KD, Schnack} it was shown that in the F-AF
delta-chain in addition to isolated magnon states there are
overlapping magnon states of a special form which are exact ground
states as well. As follows from Eq.(\ref{dispers}) the conditions
for the existence of exact magnon states do not depend on the
triangle-triangle interaction $J$ (if $J>0$) and the ground state
manifold of model (\ref{H}) on the transition line is similar to
that for the F-AF delta-chain. We note that at $J=0$ the ground
state is $6^{n}$-degenerated and the triangle-triangle interaction
$J$ lifts this degeneracy but a part of $6^{n}$ states remains
degenerated with zero energy. The number of such ground states in
each spin sector can be obtained using the results of
Refs.\cite{DK, KD, Schnack} taking into account minor corrections.
We will not give details of these calculations and present the
final results only.

For the model with open boundary condition (OBC) for any value of
$\Delta _{1}$ the number of the ground state $W_{k}(n)$ in the
spin sectors $S^{z}=S_{\max }-k$ and $S^{z}=-S_{\max }+k$ is
\begin{eqnarray}
W_{k}(n) &=&\sum_{m=0}^{k}C_{n}^{m},\qquad 0\leq k\leq n  \nonumber \\
W_{k}(n) &=&2^{n},\qquad n<k\leq \frac{3n}{2}
\end{eqnarray}%
where $C_{n}^{m}=n!/m!(n-m)!$ are binomial coefficients.

The total number of the ground states $G(n)$ is%
\begin{equation}
G(n)=\sum_{k}W_{k}(n)=(2n+1)2^{n}
\end{equation}

The number of the ground states for the model with periodic boundary
conditions (PBC) is given by%
\begin{eqnarray}
W_{k}(n) &=&C_{n}^{k},\qquad 0\leq k\leq \frac{n}{2}  \label{w(k)} \\
W_{k}(n) &=&C_{n}^{n/2},\qquad \frac{n}{2}<k\leq \frac{3n}{2}  \nonumber
\end{eqnarray}

The total number of the ground states is
\begin{equation}
G(n)=2^{n}+2nC_{n}^{n/2}  \label{G}
\end{equation}

Eq.(\ref{w(k)}) is valid for all $0<\Delta _{1}<\infty $ except
two special cases, $\Delta _{1}=\frac{1}{\sqrt{2}}$ and $\Delta
_{1}=\frac{1}{2}$. For the model with $\Delta
_{1}=\frac{1}{\sqrt{2}}$ the number $W_{k}(n)$ is
\begin{eqnarray}
W_{k}(n) &=&C_{n}^{0}+C_{n}^{2}+\ldots +C_{n}^{k},\qquad \text{even }k\leq n
\nonumber \\
W_{k}(n) &=&C_{n}^{1}+C_{n}^{3}+\ldots +C_{n}^{k},\qquad \text{odd }k\leq n-1
\nonumber \\
W_{k}(n) &=&2^{n-1},\qquad n<k\leq \frac{3n}{2}
\end{eqnarray}

The total number of the ground state for this special case is%
\begin{equation}
G(n)=2^{n}(n+1)
\end{equation}

The number of the ground state for the case $\Delta _{1}=\frac{1}{2}$ is
\begin{equation}
W_{k}(n)=C_{n}^{k}+b_{n,k}
\end{equation}%
where%
\begin{eqnarray}
b_{n,k} &=&C_{n}^{k-3},\qquad 3\leq k\leq \frac{n}{2}  \nonumber \\
b_{n,k} &=&2C_{n}^{n/2-3}-C_{n}^{n-k-3},\qquad \frac{n}{2}+1\leq k\leq n-3
\nonumber \\
b_{n,k} &=&2C_{n}^{n/2-3},\qquad n-2\leq k\leq \frac{3n}{2}
\end{eqnarray}

The total number of the ground states in this case is%
\begin{equation}
G(n)=2^{n}+2nC_{n}^{n/2}+6(n-2)C_{n}^{n/2-3}
\end{equation}

We note that the ground state degeneracies $W_{k}(n)$ and $G(n)$
on the transition line do not depend on $J$ including the case
$J\to \infty $, when spin-$\frac{1}{2}$ triangle chain model
reduces to the F-AF delta-chain with the basal spins $s_{2}=1$ and
the apical spins $s_{1}=\frac{1}{2}$.

Though the numbers of the ground states for the models with OBC
and PBC are different for finite chains, they are the same (with
an exponential accuracy) in the thermodynamic limit $n\to \infty $
and $G(n)\simeq 2^{n}$. As a result the residual entropy is
\begin{equation}
\mathcal{S}=\frac{\ln G}{N}=\frac{1}{3}\ln 2  \label{entropy}
\end{equation}

We note also that the ground state degeneracy at $J>0$ is
$G(n)\simeq 2^{n}$ in comparison with $G(n)=6^{n}$ at $J=0$.

The spin-$\frac{1}{2}$ triangle chain at the transition point can
be extended to $s_{1}$, $s_{2}$-triangle chain in which the apical
and the basal spins are $s_{1}$ and $s_{2}$ correspondingly and
the interaction parameter $\alpha =\frac{1}{2\Delta
_{1}}\frac{s_{1}}{s_{2}}$. In particular, this parameter is
$\alpha =\frac{1}{4}$ for the delta-chain with
$s_{1}=\frac{1}{2}$, $s_{2}=1$ which corresponds to the limit
$J\to\infty $ of the $s=\frac{1}{2}$ triangle chain. The ground
state degeneracy of the $s_{1}$, $s_{2}$-triangle chain can be
obtained in the same manner as for the case
$s_{1}=s_{2}=\frac{1}{2}$.

It is interesting to consider the influence of the exponentially
large ground state manifold on low-temperature thermodynamics of
the model in the magnetic field $h$. The contribution of the
degenerate ground states to the partition function $Z_{gs}$ can be
calculated analytically and $Z_{gs}$ has a form (we consider the
periodic model)
\begin{equation}
Z_{gs}=2\sum_{k=0}^{n/2}C_{n}^{k}\cosh \left[
(\frac{3n}{2}-k)\frac{h}{T}\right]
+2C_{n}^{n/2}\sum_{k=n/2}^{3n/2}\cosh \left[ (\frac{3n}{2}-k)\frac{h}{T}%
\right]  \label{Z}
\end{equation}

We note that $Z_{gs}$ and all thermodynamics depend on an
universal variable $x=h/T$. Of course, the contribution of excited
states destroys this universality (see below).

The magnetization $M$ and the susceptibility $\chi $ are given by
\begin{equation}
M=T\frac{d\ln Z}{dh},\qquad \chi =T\frac{d^{2}\ln Z}{dh^{2}}  \label{Z1}
\end{equation}

The analysis of Eqs.(\ref{Z}) and (\ref{Z1}) shows that the
magnetization and the susceptibility depend on the relation
between $x$ and $N$. It can be shown that $M$ at $x\gg
\frac{1}{N}$ has a form
\begin{equation}
M=\frac{3+e^{-x}}{6+6e^{-x}}N  \label{m3}
\end{equation}

\begin{figure}[tbp]
\includegraphics[width=4in,angle=0]{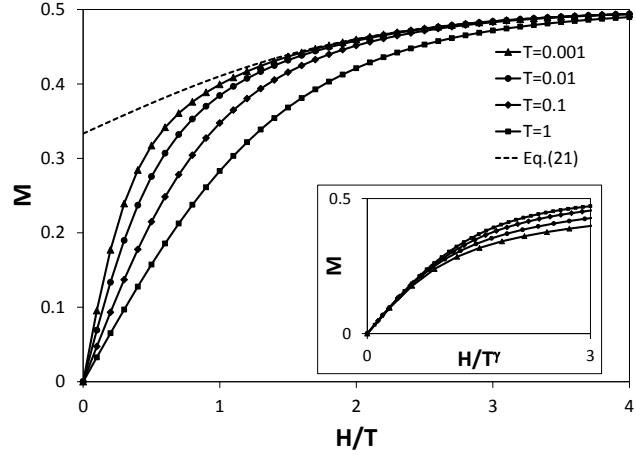}
\caption{Magnetization as a function of $h/T$ for isotropic model
(\ref{H}) with $J=1$ and $n=6$ at several values of temperature.
The inset shows the magnetization vs scaled magnetic field
$h/T^{\gamma}$.} \label{M(H)_isotropic_J1}
\end{figure}

As follows from Eq.(\ref{m3}) the magnetization per spin
$m=\frac{M}{N}$ at $\frac{h}{T}\to 0$ tends to $m=\frac{1}{3}$.
Therefore, the spin triangle chain model on the transition line is
magnetically ordered at zero temperature and the magnetization
undergoes a jump from $m=\frac{1}{2}$ in the ferromagnetic phase
to $m=\frac{1}{3}$ on the transition line. The magnetization
$m(x)$ given by Eq.(\ref{m3}) is shown in
Fig.\ref{M(H)_isotropic_J1} together with that obtained by the
exact diagonalization (ED) calculations for the temperatures from
$T=10^{-3}$ to $T=1$. As can be seen from
Fig.\ref{M(H)_isotropic_J1} the magnetization $m$ is, in fact, the
function of both $x$ and $T$.

The magnetization $M$ given by Eqs.(\ref{Z}) and (\ref{Z1}) at
$x\ll \frac{1}{N} $ is
\begin{equation}
M=d_{n}x,\qquad d_{n}=\frac{2^{n}\left( n^{2}+\frac{n}{4}\right)
+C_{n}^{n/2}(\frac{2}{3}n^{3}+n^{2}+\frac{n}{3})}{2^{n}+2nC_{n}^{n/2}}
\label{m2}
\end{equation}%
and at $N\gg 1$ (but $x\ll \frac{1}{N})$ the magnetization is
\begin{equation}
M=\frac{N^{2}x}{27}  \label{m1}
\end{equation}

The uniform magnetic susceptibility $\chi $ at $T\to 0$ can be
found from Eq.(\ref{m2}) and it is $\chi =d_{n}/T$. The quantity
$\chi T/N$ at $T\to 0$ is finite and increases with $N$ as $\chi
T/N=0.823(N=12), 1.1038(N=18), 1.3764(N=24), 1.6436(N=30)$.
%
%
For $N\gg 1$, $\chi T/N=N/27$, i.e. it is proportional to $N$
instead of to be a constant. It means that $\chi T/N$ diverges at
$T\to 0$ in the thermodynamic limit and the behavior of $\chi $ at
large $N$ and small $T$ is described by the finite-size scaling
function $\chi =T^{-\gamma }f(NT^{\gamma -1})$ and $\frac{\chi
}{N}\sim T^{-\gamma }$ in the thermodynamic limit. Generally, the
critical exponent $\gamma $ depends on $\Delta _{1}$ and $J$. We
estimate the value of $\gamma $ for the isotropic model ($\Delta
_{1}=1$) and for some values of $J$ using the results of the ED
calculations of the magnetization for the chain with $n=6$
triangles ($N=18$). An analysis of data on inset in
Fig.\ref{M(H)_isotropic_J1} shows that the magnetization at small
values of $h$ and $T$ corresponds to the scaling variable
$y=h/T^{\gamma }$ with $\gamma \simeq 1.15$.
%

\begin{figure}[tbp]
\includegraphics[width=4in,angle=0]{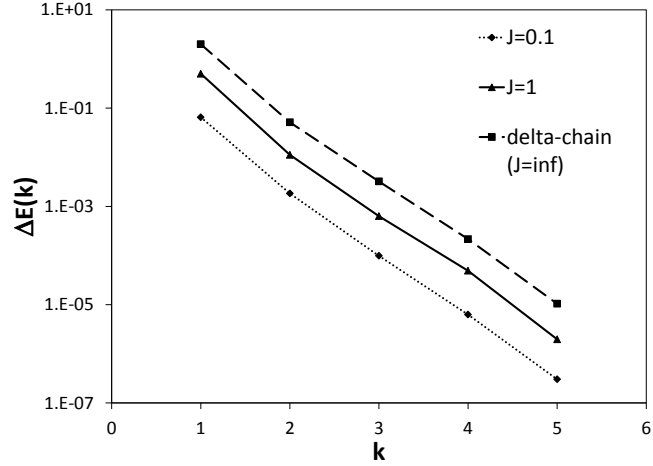}
\caption{The gap for $k$-magnon states, $\Delta E(k)$, for
isotropic model (\ref{H}) with $n=10$ and $J=0.1, 1$ and
$\Delta$-chain with basal spins $s=1$, which corresponds to the
limit $J\to\infty$.} \label{DE(k)n10}
\end{figure}

We note that the partition function $Z_{gs}$ in Eq.(\ref{Z}) at
$h=0$ does not depend on the temperature and the thermodynamics at
$h=0$ is determined by the contribution of the excited states
only. Therefore, it is necessary to study the properties of the
spectrum of excitations (with a particular focus on the isotropic
case, $\Delta _{1}=1$). One of these points is related to the
value of the lowest excitation in the spin sector $S^{z}=S_{\max
}-k$ (the gap for $k$-magnon states). The gap $\Delta E$ for $k=1$
can be found from Eq.(\ref{k=1 excitation}). In particular, for
the isotropic model $\Delta E=\frac{2}{3}J$ for $J\ll 1$ and
$\Delta E=\frac{1}{2}$ for $J=1$ and $\Delta E=2$ for
$J\to\infty$. The gaps decrease rapidly with increasing $k$ as it
can be seen from Fig.\ref{DE(k)n10} where the gaps for the
periodic isotropic chain for $n=10$ ($N=30$) are presented for
some values of $J$ including the case $J=\infty $. On the base of
these results it can be concluded that the gap becomes extremely
(exponentially) small for $k\gg 1$. Another important observation
that follows from Fig.\ref{DE(k)n10} is that the behavior of
$\Delta E(k)$ for different values of $J$ are very similar, and
coincide with very high accuracy after the corresponding scaling
of energy. The latter fact has important consequences that will be
discussed below.

%
%
%

\begin{figure}[tbp]
\includegraphics[width=4in,angle=0]{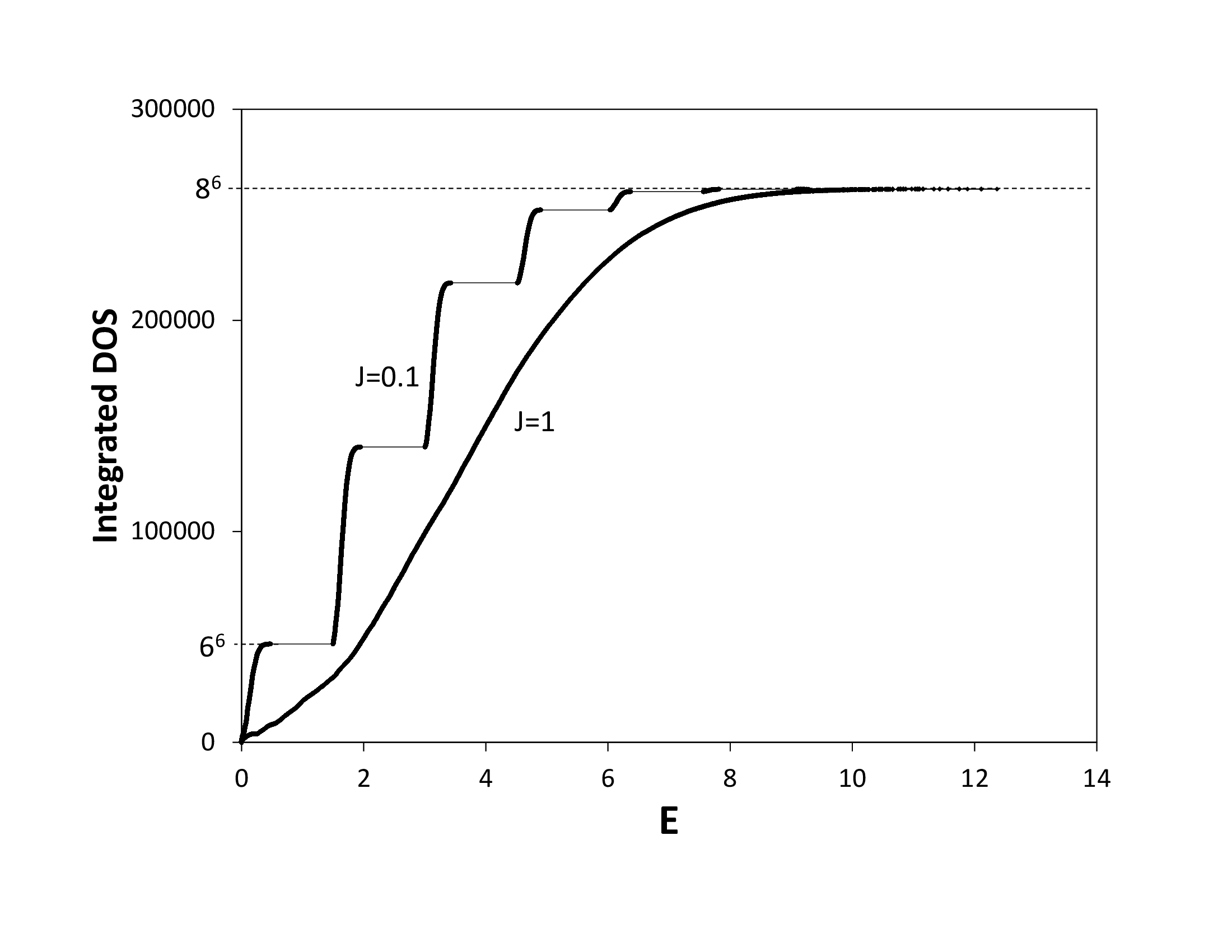}
\caption{Integrated density of states of the isotropic model
(\ref{H}) for $J=0.1; 1$ and $n=6$.} \label{L6_isotropic_J01_J1}
\end{figure}

\begin{figure}[tbp]
\includegraphics[width=4in,angle=0]{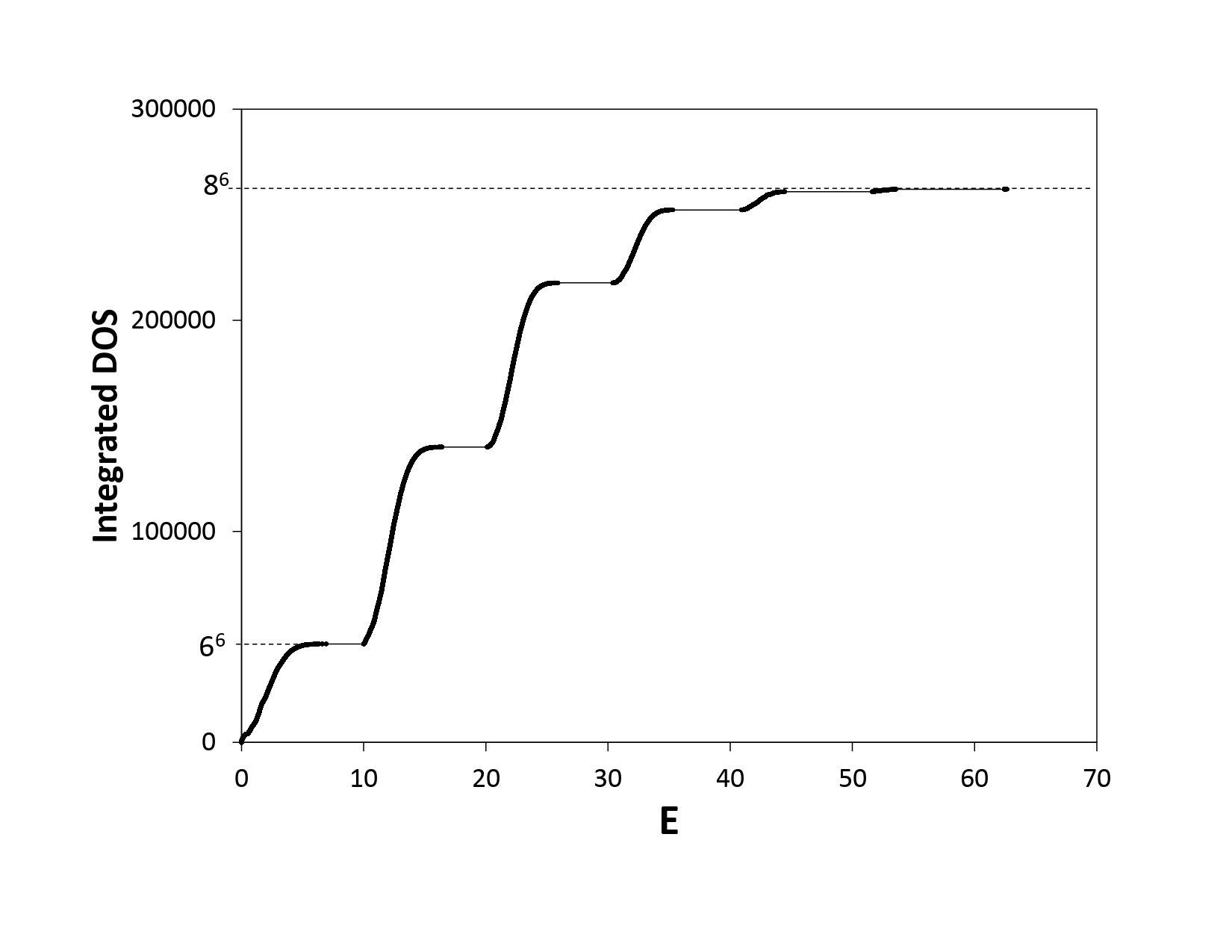}
\caption{Integrated density of states of the isotropic model
(\ref{H}) for $J=10$ and $n=6$.} \label{L6_isotropic_J10}
\end{figure}

\begin{figure}[tbp]
\includegraphics[width=4in,angle=0]{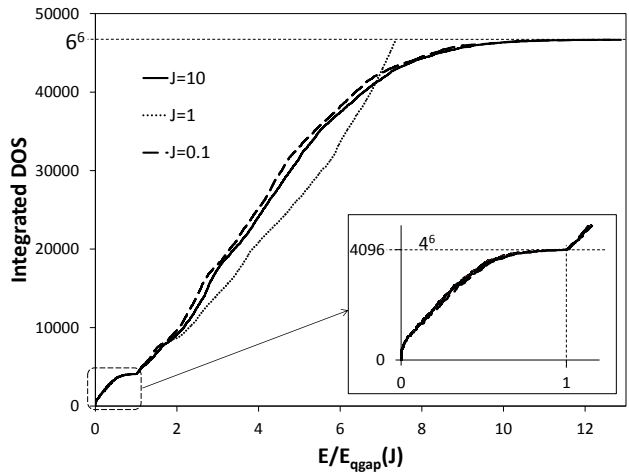}
\caption{Low-energy part of the integrated density of states of
the isotropic model (\ref{H}) plotted against the energy
normalized by $E_{qgap}$ for $J=0.1; 1; 10$ and $n=6$. The inset
shows the lowest energy part of the plot.} \label{DOS
isotropic_cusp}
\end{figure}

\begin{figure}[tbp]
\includegraphics[width=4in,angle=0]{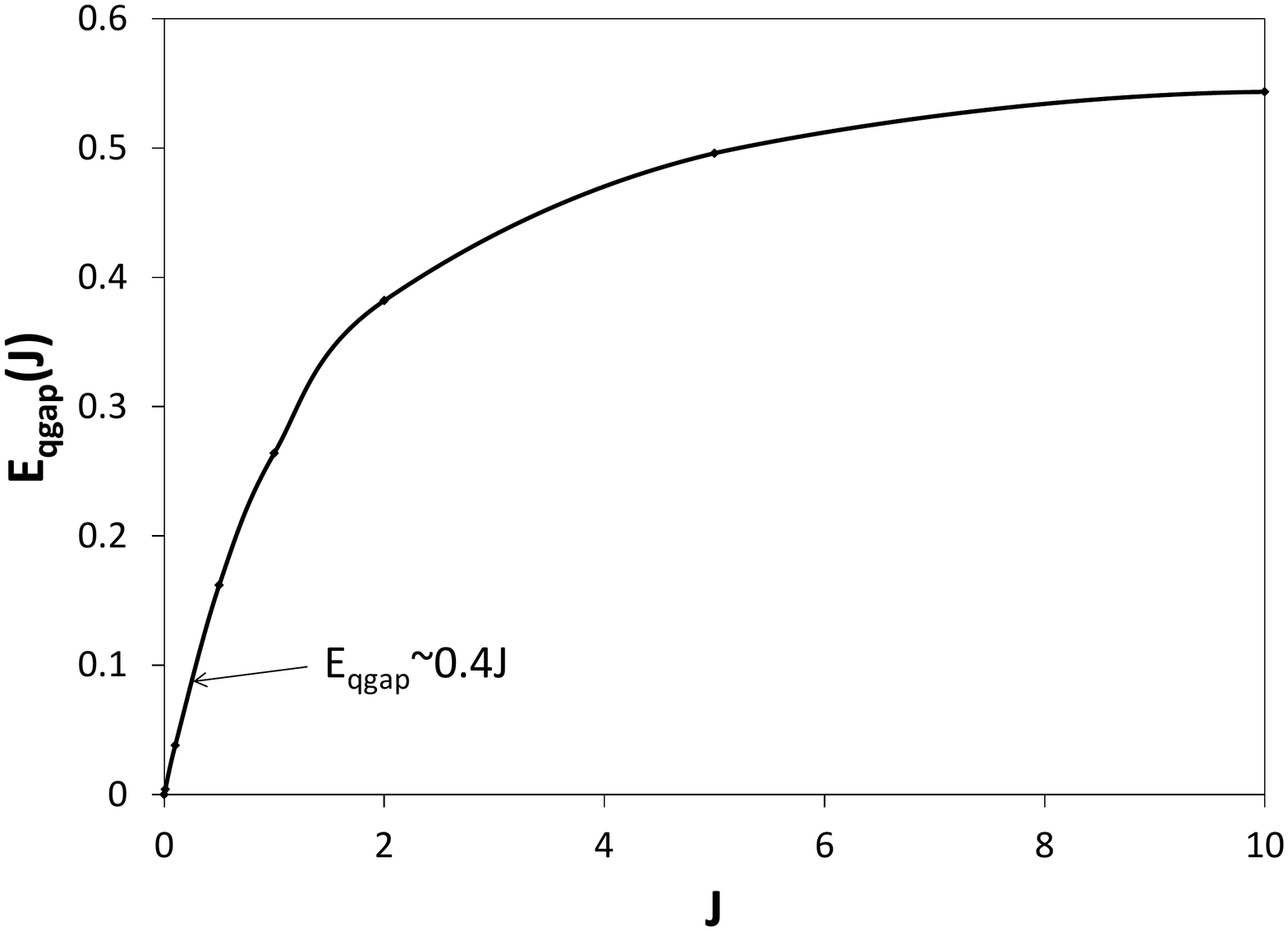}
\caption{Dependence of the energy of the $4^n$-th level that
separates the lowest-energy scale on $J$ for the isotropic model
(\ref{H}).} \label{E_cusp(J)}
\end{figure}

A significant characteristic of the spectrum is the distribution
of the energy levels (density of states). As an example, we
consider this distribution for the isotropic model. At $J=0$ the
spectrum consists of $(n+1)$ isolated levels with energies
$E_{0}=0, E_{1}=\frac{3}{2}, E_{2}=3,\ldots E_{n}=\frac{3n}{2}$.
The total number of states is $8^{n}$. The lowest level with
$E_{0}=0$ contains $6^{n}$ states and the level with
$E_{k}=\frac{3k}{2}$ contains $C_{n}^{k}6^{n}/3^{k}$ states. The
neighboring levels are separated from each other by the gap
$\Delta E=\frac{3}{2}$. At $J>0$ each level splits and transforms
to the band and a width of the band depends on $J$. In order to
find out the dependence of the spectrum properties on $J$ we
calculate the integrated density of states for the isotropic model
with $n=6$ and $J=0.1; 1; 10$ which are shown in
Figs.\ref{L6_isotropic_J01_J1},\ref{L6_isotropic_J10}. As it is
seen from these Figures the spectrum for small and large $J$ has a
step-like structure with gaps between neighboring bands but the
spectrum is gapless for $J=1$. Our numerical calculations show
that the gap $\Delta E$ between the lowest band and the next one
decreases with increasing $J$ from the value $\Delta
E=\frac{3}{2}$ at $J=0$ to zero at $J\simeq 1 $. But then $\Delta
E$ grows again and the gap becomes $\Delta E\simeq J$ for $J\gg
1$. The structure of the lowest band containing $6^{6}$ states is
of particular interest because it determines the low-temperature
thermodynamics. Corresponding integrated density of states is
shown in Fig.\ref{DOS isotropic_cusp}. As it can be seen from
Fig.\ref{DOS isotropic_cusp} in this band there is a low-energy
region separated by a quasi-gap (a cusp in the density of states)
from the region above the cusp as it is shown in Fig.\ref{DOS
isotropic_cusp} (inset). A notable feature of the spectrum is the
fact that the number of states in the low-energy part is $4^{6}$.
We note that a similar spectrum takes place for the model with
$n=4$, where the number of levels below the quasi-gap is $4^{4}$.
Based on these facts, we assume that such specific structure of
the spectrum is a general property of the model and the low-energy
part consists of $4^{n}$ levels below the quasi-gap. This
low-energy region also includes $\sim 2^{n}$ ground states with
zero energy.

The position of the quasi-gap depends on $J$ and the results of
the calculations of this dependence are shown in
Fig.\ref{E_cusp(J)} for the chain with $n=6$. At small values of
$J$ the quasi-gap energy is $\sim 0.4J$ and it tends to $\sim 0.6$
for large $J$. A remarkable feature of the spectrum is that the
integrated density of states in the energy region below the cusp
is an universal function of the normalized energy $E/E_{qgap}(J)$
for any $J$ as it is shown in Fig.\ref{DOS isotropic_cusp}
(inset). Similar to this the densities of states in the band of
$6^{6}$ states as functions of $E/E_{qgap}(J)$ are very close to
each other for small and large $J$ as it is shown in Fig.\ref{DOS
isotropic_cusp}. Another remarkable peculiarity of the spectrum is
related to a ratio of a width of the lowest band of $6^{6}$ states
to the quasi-gap energy $E_{qgap}(J)$. It is almost the same for
both small and large $J$ (it is $\sim 13$) and it is $\sim 7$ for
$J\simeq 1$. We note that these unusual feature of the spectrum as
well as the dependence of the quasi-gap energy on $J$ takes place
for the model with any $n$ as it follows, for example, from the
temperature dependence of the specific heat.

\begin{figure}[tbp]
\includegraphics[width=4in,angle=0]{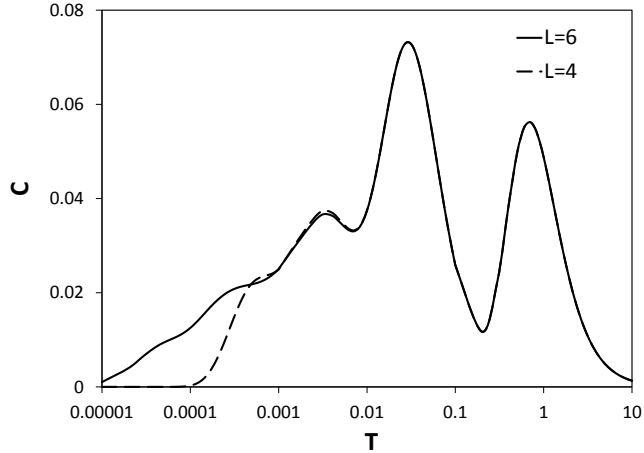}
\caption{Dependence of the specific heat on the temperature of the
isotropic Hamiltonian (\ref{H}) with $J=0.1$ and $n=4$ (dashed
line) and $n=6$ (solid line).} \label{C(T)_J01_L4L6}
\end{figure}

This specific structure of the spectrum determines the
low-temperature thermodynamics. In Fig.\ref{C(T)_J01_L4L6} we
represent the data for the temperature dependence (in logarithmic
temperature scale) of the specific heat $C(T)$ obtained by the ED
calculation for the isotropic model with $n=4$ and $n=6$ for
$J=0.1$. As it can be seen from Fig.\ref{C(T)_J01_L4L6} the data
for $n=4$ and $n=6$ deviate from each other for $T<10^{-4}$ but
are close for $T>10^{-3}$. It indicates that the data for $n=6$
describe the thermodynamic limit at $T>10^{-3}$.

\begin{figure}[tbp]
\includegraphics[width=4in,angle=0]{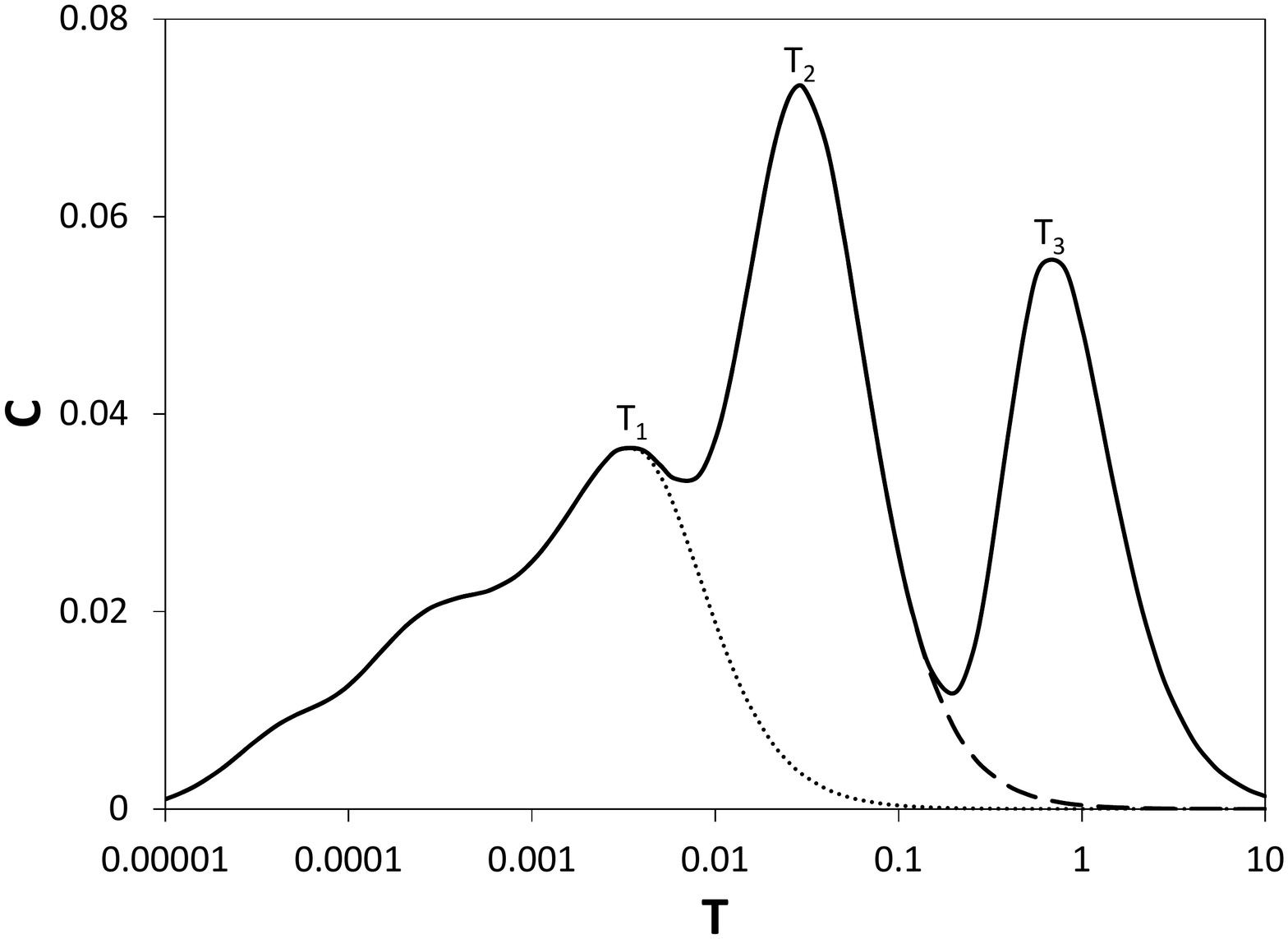}
\caption{Dependence of the specific heat on the temperature of the
isotropic Hamiltonian (\ref{H}) with $J=0.1$ and $n=6$ (solid
line). The specific heat calculated on truncated partition
function containing the lowest $4^n$ (dotted line) and $6^n$
(dashed line) states.} \label{C(T)_J01_truncated}
\end{figure}

\begin{figure}[tbp]
\includegraphics[width=4in,angle=0]{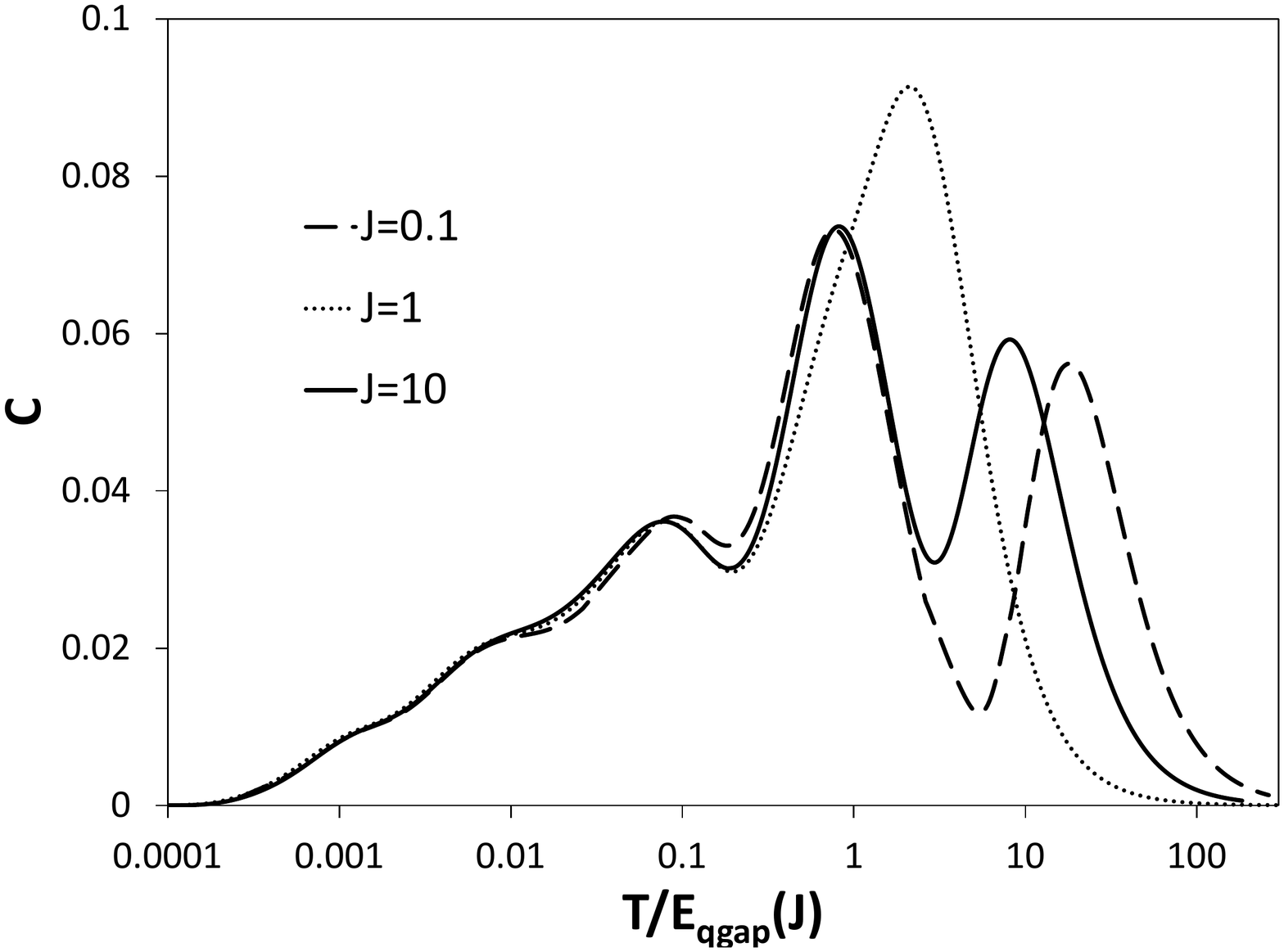}
\caption{Dependence of the specific heat on the temperature of the
isotropic Hamiltonian (\ref{H}) with $J=0.1; 1; 10$ and $n=6$.}
\label{C(T)_J}
\end{figure}

The specific heat in Fig.\ref{C(T)_J01_truncated} for $J=0.1$ is
characterized by the existence of two pronounced maxima at
$T\simeq 0.03$ and $T\simeq 1$. In addition to these maxima there
is a small low-temperature peak at $T\simeq 0.003$. We propose
that the left broad maximum at $T=0.03$ is determined by $6^{n}$
states of the lowest band and the right maximum at $T\simeq 1$ is
related to other high-energy states. Let us denote positions of
three maxima as $T_{1}$, $T_{2}$ and $T_{3}$ from left to right,
correspondingly. When $J$ is small, then $T_{2}<T_{3}$. When $J$
increases up to $J\simeq 1$ both broad maxima merge ($T_{2}\simeq
T_{3}$) and the spectrum becomes gapless. After that they separate
again when $J$ increases. At $J\to\infty$ both $T_{1}$ and $T_{2}$
are finite ($T_{1},T_{2}\simeq 1 $) while $T_{3}\to\infty$ as a
consequence of removing of the part of high-energy states above
the gap to an infinity together with the gap which is $\Delta
E\simeq J$. Therefore, in the limit $J\to\infty $ there are two
maxima and $C(T)$ behave as that for the delta-chain. On the other
hand, in the limit $J\to 0$ both $T_{1}$ and $T_{2}$ tend to zero
and $T_{3}\simeq 1$. As a result the specific heat for $J=0$ has
one maximum at $T\simeq 1$.

In Fig.\ref{C(T)_J} we represent the temperature dependence of the
specific heat for the isotropic model with $n=6$ and $J=0.1$; $1$;
$10$. Taking to account an universality of the density of
low-energy states as a function of $E/E_{qgap}(J)$ we renormalize
the temperature scale as $y=T/E_{qgap}(J) $. Then the results for
small and large $J$ are close with each other for different $J$ at
$y\leq 1$. To confirm this relation we calculate the contribution
of each part of the spectrum to the specific heat. At first, we
select the part containing $6^{6}$ states and remove the
high-energy states above the gap. The contribution of this part is
responsible for the maxima of $C(T)$ at $y\simeq 1$. At second, we
select $4^{6}$ low-energy states of the spectrum below the cusp
and remove other states. The contributions of these parts are
shown in Fig.\ref{C(T)_J01_truncated} and they reproduce very well
corresponding maxima of $C(T)$.

One remark is concerned to the spectrum of model (\ref{H}) in the
limit $J\to\infty $. In this limit model (\ref{H}) reduces to the
delta-chain with the apical spin $s_{1}=\frac{1}{2}$ and the basal
spin $s_{2}=1$. The total number states in such model is $6^{n}$
in comparison with the number $8^{n}$ for the spin triangle chain
with finite $J$. But the spectrum of such delta-chain has one band
consisting of two parts and the low-energy region contains $4^{n}$
states which is separated by the cusp from the high-energy region,
i.e. the structure of the spectrum is the same as for the lowest
band of the spin-$\frac{1}{2}$ triangle chain with finite $J$. The
specific heat of such delta-chain has two maxima in accordance
with above mentioned transformation of the spectrum at
$J\to\infty$.

\begin{figure}[tbp]
\includegraphics[width=4in,angle=0]{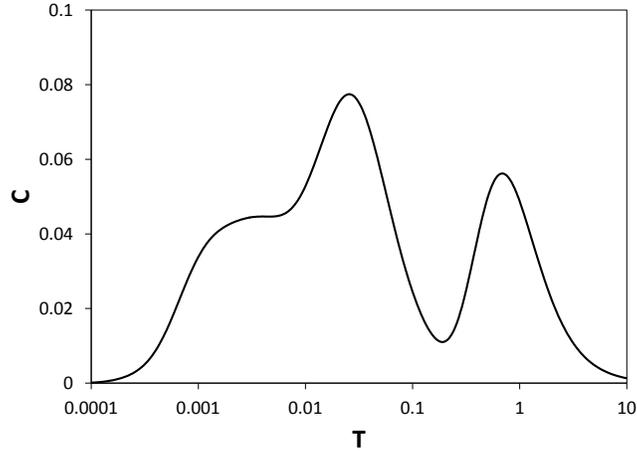}
\caption{Dependence of the specific heat on the temperature of the
anisotropic Hamiltonian (\ref{H}) with $\Delta_1 =0.5$ and $n=6$.}
\label{C(T)_D05_J01}
\end{figure}

A similar temperature dependence of the specific heat is realized
for other values of $\Delta_{1}<1$ as is shown in
Fig.\ref{C(T)_D05_J01} with shoulder instead of the peak at low
temperature. The temperature dependence of the entropy per site
$s(T)$ demonstrates a stair-step behavior between limits
$s_{0}=\frac{1}{3}\ln 2$ at $T\to 0$ and $s=\ln 2$ at $T\to
\infty$.

The $s=\frac{1}{2}$ triangle chain on the transition line between
the ferromagnetic and other ground states can be extended to the
triangle chain in which the apical and the basal spins are $s_{1}$
and $s_{2}$, correspondingly. Such model is the one-parametric as
well and we can take as this parameter the anisotropy $\Delta
_{1}$. Then the basal--basal interaction parameters $\alpha $ and
$\Delta _{2}$ are $\alpha =\frac{1}{2\Delta
_{1}}\frac{s_{1}}{s_{2}}$ and $\Delta _{2}=2\Delta _{1}^{2}-1$.
The ground state degeneracy of the $s_{1},s_{2}$-triangle chain
can be obtained from Eq.(\ref{w(k)}) with $S_{\max
}=(2s_{2}+s_{1})n$ instead of $\frac{3n}{2}$. The ground state
degeneracy does not depend on $J$. In the limit $J\to \infty $
spin- $s_{1}$, $s_{2}$ triangle chain reduces to the delta-chain
with the apical spin $s_{1}$ and the basal spin $2s_{2}$. In
particular, the $s=\frac{1}{2}$ triangle chain becomes the F-AF
delta-chain with $s_{1}=\frac{1}{2},s_{2}=1$ and with the
interaction parameter $\alpha =\frac{1}{4\Delta _{1}}$.

It is interesting to consider the distribution of the energy
levels in the spectrum of the spin- $s_{1}$, $s_{2}$ triangle
chain. The total number of states is
($2s_{2}+1)^{2n}$($2s_{1}+1)^{n}$. The distribution has the band
structure and the lowest band is separated by the gap $\Delta
E\simeq 1$ from the high-energy region. The number of states below
the gap is $4^{n}(s_{1}+2s_{2})^{n}$. The numerical calculations
of the integrated density of states for the spin triangle chains
for some values of $s_{1}$ and $s_{2}$ show that the lowest band
in the model with the apical spin $s_{1}=\frac{1}{2}$ and the
basal spin $s_{2}$ has low-energy region separated by the cusp
from the region above the cusp. The number of states in the region
below the cusp is $(4s_{2}+2)^{n}$ and is the same for any $J$.
This region includes the ground state manifold with zero energy.
We note that in the limit $J\to \infty $ in which the
$s_{1},s_{2}$-triangle chain reduces to the
$s_{1},2s_{2}$-delta-chain there is one band in the spectrum with
$2^{n}(4s_{2}+1)^{n}$ states and there is low-energy region below
the cusp and the number of states in this region is
$(2s_{2}+2)^{n}$ as for the triangle chain with finite $J$. It is
interesting to compare the properties of the spectrum of
spin-$\frac{1}{2}$ triangle chain with those for the F-AF
spin-$\frac{1}{2}$ delta-chain. The spectrum of the latter
contains $6^{n}$ states and there is a cusp in the density of
states. The number of states below the cusp is $3^{n}$. As was
shown in \cite{DK} such structure of the spectrum leads to the
temperature dependence $C(T)$ with two maxima in contrast with
that for spin-$\frac{1}{2}$ triangle chain in which $C(T)$ has
three peaks.

\section{The spin-$\frac{1}{2}$ triangle chain with strong anisotropy of the
interactions}

In this Section we consider the spin triangle model in the
transition line in the strong anisotropic limit, $\Delta _{1}\gg
1$. It is convenient to normalize the Hamiltonian (\ref{H1}) as
$H_{i}/\Delta _{1}$ and to represent it in a form
\begin{eqnarray}
\tilde{H}_{i}
&=&-2g\sum_{m=1,2}(s_{i,m}^{x}s_{i,m+1}^{x}+s_{i,m}^{y}s_{i,m+1}^{y})-%
\sum_{m=1,2}(s_{i,m}^{z}s_{i,m+1}^{z}-\frac{1}{4})  \nonumber \\
&&+2g^{2}(s_{i,1}^{x}s_{i,3}^{x}+s_{i,1}^{y}s_{i,3}^{y})+(1-2g^{2})(s_{i,1}^{z}s_{i,3}^{z}-%
\frac{1}{4})
\end{eqnarray}%
where $g=\frac{1}{2\Delta _{1}}$.

The Hamiltonian of the model is%
\begin{equation}
H=\sum_{i=1}^{n}\tilde{H}_{i}+\sum_{i=1}^{n}V_{i,i+1}  \label{Hg}
\end{equation}

At first we consider the limit $g=0$. At $g=0$ the model
(\ref{Hg}) reduces to the Ising triangle model with the Heisenberg
ferromagnetic interaction between them. The Ising basal-apical and
the basal-basal interactions in each triangle are $-1$ and $+1$,
correspondingly (in the limit $J\to \infty $ they are $-1$ and
$+\frac{1}{2}$).

Of course, the spectrum of such Ising-Heisenberg Hamiltonian
depends on $J$ and in the limit $J\to \infty $ even the total
number of states is $6^{n}$ rather than $8^{n}$ for $J>0$. But
here we are interested in the low-energy part of the spectrum
containing just $6^{n}$ states and the numbers of the $k $-magnon
states in this part of the spectrum are the same for any $J$
including the limiting case $J=\infty $.

A most simple way to determine these numbers is just to consider
this limiting case. As we note above in the limit $J=\infty $
model (\ref{Hg}) for $g=0$ reduces to the Ising delta-chain with
the apical spin $s=\frac{1}{2}$ and the basal spin $\tau =1$. The
Hamiltonian of this Ising model has a form
\begin{equation}
H_{\mathrm{Ising}}=\sum_{i=1}^{n}[-s_{2i}^{z}(\tau _{2i-1}^{z}+\tau
_{2i+1}^{z}-1)+\frac{1}{2}(\tau _{2i-1}^{z}\tau
_{2i+1}^{z}-1)-h(s_{2i}^{z}+\tau _{2i-1}^{z})]  \label{Ising}
\end{equation}%
where $s_{2i}^{z}=(\frac{1}{2},-\frac{1}{2})$, $\tau _{2i-1}^{z}=(1,0,-1)$
and $h$ is the magnetic field.

The partition function of the Ising model (\ref{Ising}) can be
obtained using the transfer-matrix method and the eigenvalues of
the transfer matrix satisfy the equation
\begin{equation}
\lambda ^{3}-\lambda ^{2}(1+y^{2})-\lambda
y(1+y+y^{2}+y^{3}+y^{4})-y^{4}(1+y)=0  \label{eigen}
\end{equation}%
where $y=\exp (-\frac{h}{T})$.

The partition function $Z_{\mathrm{Ising}}$ is%
\begin{equation}
Z_{I\sin g}=\lambda _{1}^{n}+\lambda _{2}^{n}+\lambda _{3}^{n}  \label{Zi}
\end{equation}

The ground state energy of (\ref{Ising}) is zero and the ground
state degeneracy in the spin sector $S^{z}=(\frac{3n}{2}-k)$
($0\leq k\leq 3n$) can be found as coefficients in the expansion
of $Z_{\mathrm{Ising}}$ in powers of $\exp (-\frac{h}{T})$. The
first terms of an expansion of $\lambda $ on $y$ are
\begin{eqnarray}
\lambda _{1} &=&1+y+2y^{3}-y^{4}+3y^{5}-8y^{6}+\ldots  \nonumber \\
\lambda _{2} &=&-y+y^{4}-3y^{5}+\ldots  \nonumber \\
\lambda _{3} &=&-y^{3}+y^{6}+\ldots
\end{eqnarray}

The number of the ground states of model (\ref{Ising}) $W_{k}(n)$
in the spin sector $S^{z}=S_{\max }-k$ is
\begin{equation}
W_{k}(n)=C_{n}^{k}+2(k-2)C_{n}^{k-2}-(k-3)C_{n}^{k-3}
+3(k-4)C_{n}^{k-4}-8(k-5)C_{n}^{k-5}+\ldots
\end{equation}

In particular, the number of the ground states in the spin sector
$S^{z}=S_{\max }-k$ with $k\leq 4$ is
\begin{equation}
W_{1}(n)=n,\qquad W_{2}(n)=C_{n}^{2},\qquad W_{3}(n)=C_{n}^{3}+2n,\qquad
W_{4}(n)=C_{n}^{4}+4nC_{n}^{2}-n  \label{w1(k)}
\end{equation}

The total number of the ground states of model (\ref{Ising})
$G_{0}(n)$ is given by Eq.(\ref{Zi}) with $\lambda _{i}$ at $y=1$.
The solutions of Eq.(\ref{eigen}) at $y=1$ are
\begin{equation}
\lambda _{1}=\frac{3+\sqrt{17}}{2},\qquad \lambda _{2}=\frac{3-\sqrt{17}}{2}%
,\qquad \lambda _{3}=-1
\end{equation}%
and $G_{0}(n)$ is%
\begin{equation}
G_{0}(n)=\lambda _{1}^{n}+\lambda _{2}^{n}+\lambda _{3}^{n}
\end{equation}

For $n\gg 1$ the total number of the ground states is
$G_{0}(n)=\lambda _{1}^{n}=(3.56..)^{n}$.

At $g=0$ the ground state of the Hamiltonian (\ref{Hg}) is
$(3.56..)^{n}$-fold degenerate and at $g\neq 0$ the degeneracy is
lifted for each spin sector $S^{z}=(S_{\max }-k)$ but only partly:
some ground states remain degenerate with zero energy and the
number of such states is given by Eq.(\ref{w(k)}). But other
states move up. As a result the ground state degeneracy of model
(\ref{Hg}) at $g\neq 0$ is $G(n)\simeq 2^{n}$ in contrast with
$G_{0}(n)=(3.56..)^{n}$. At $g\ll 1$ split off levels form a set
of low-lying excitations determining the low-temperature
thermodynamics.

In the one-magnon sector ($k=1$) the number of the ground states
is $W_{1}(n)=n$ for both cases $g=0$ and $g\neq 0$. The same thing
takes place in the two-magnon sector where $W_{2}(n)=C_{n}^{2}$
for $g=0$ and $g\neq 0$. In the three-magnon sector ($k=3$) there
are $C_{n}^{3}$ ground states at $g\neq 0$ and $2n$ states
according to Eq.(\ref{w1(k)}) split off from the ground state
manifold at $g=0$ and they are bound magnon complexes. The
calculation of these complexes is rather cumbersome and we give a
final result for $k=3$. There are $n$ split off states with the
energy $E\sim g^{4}$ and $n$ states with $E\sim g^{2}$. The exact
diagonalization (ED) calculation of four-magnon complexes show
that there are $n$ split off states with $E\sim g^{8}$, then
$n(n-5)$ states with $E\sim g^{4}$ and $n(n+1)$ states with $E\sim
g^{2}$. The total number of split off states is ($4nC_{n}^{2}-n$)
in agreement with the value $W_{4}(n)$ in Eq.(\ref{w1(k)}).

On a base of the numerical calculations we conclude that the
energy of the lowest excitations at $g\ll 1$ in $k$-magnon sector
with $k\geq 3$ is $E\sim g^{4(k-2)}$, then there are excitations
with $E\sim g^{4(k-3)}, E\sim g^{4(k-4)},\ldots$ up to $E\sim
g^{4(\frac{k}{2}-1)}$, and $E\sim g^{2(k-3)}, E\sim
g^{2(k-4)},\ldots E\sim g^{2}$. Thus, the low-energy excitations
in the spin sector with $k$ magnons are divided into parts with
the energies from $E\sim g^{2}$ to $E\sim g^{4(k-2)}$.

\begin{figure}[tbp]
\includegraphics[width=4in,angle=0]{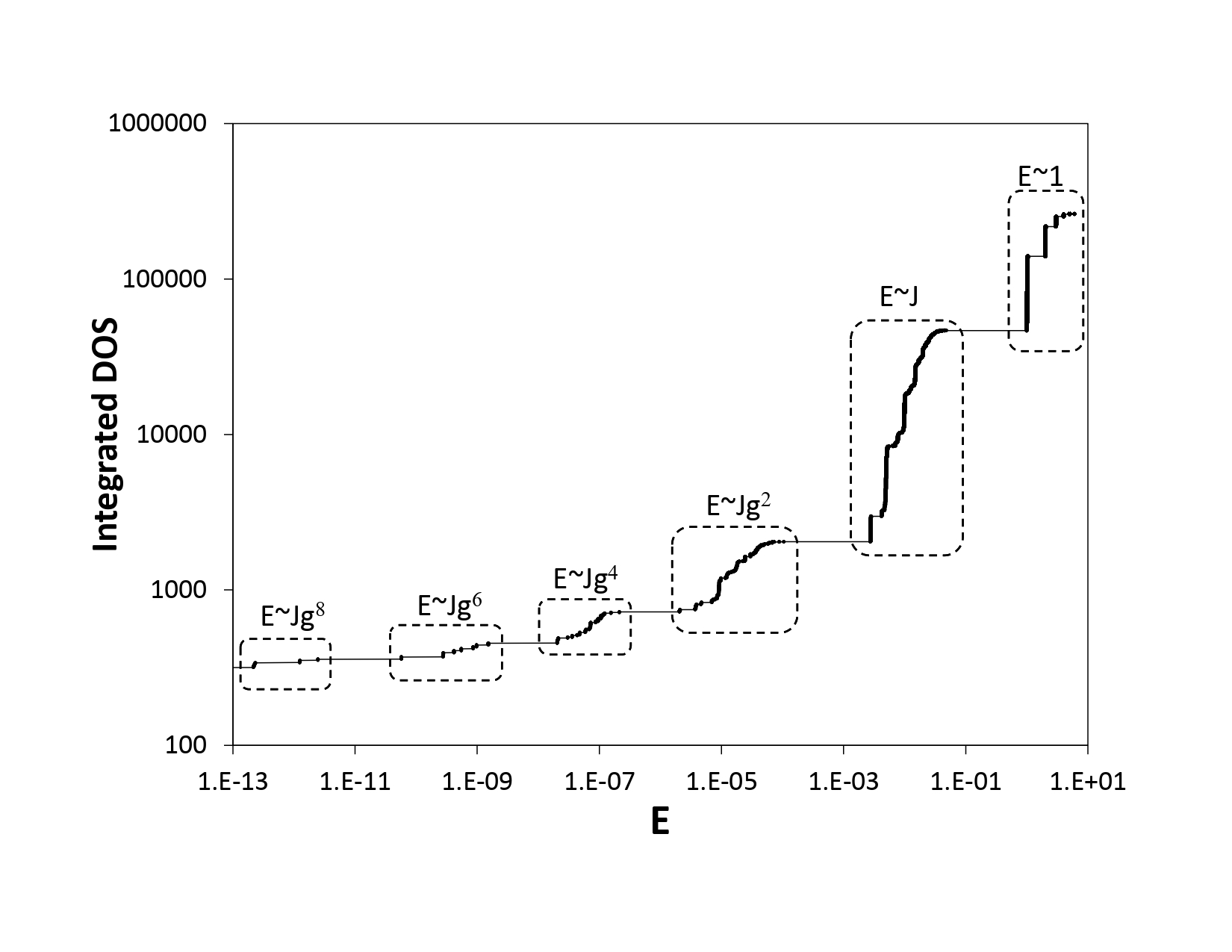}
\caption{Integrated density of states of the normalized
Hamiltonian (\ref{Hg}) with $g=0.05$, $J=0.01$ and $n=6$.}
\label{spectrum_Delta10}
\end{figure}

\begin{figure}[tbp]
\includegraphics[width=4in,angle=0]{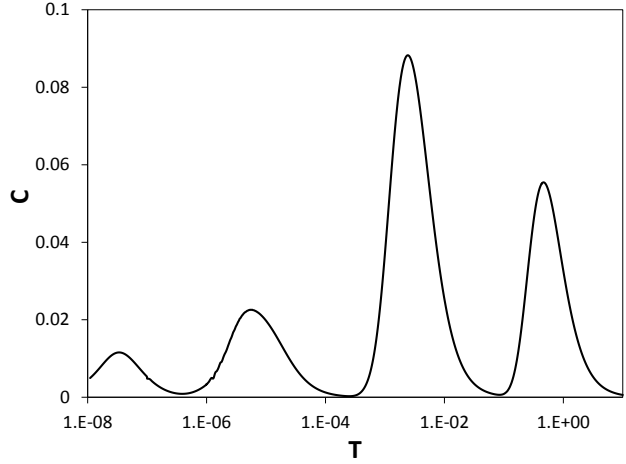}
\caption{Dependence of the specific heat on the temperature of the
normalized Hamiltonian (\ref{Hg}) with $g=0.05$, $J=0.01$ and
$n=6$.} \label{C(T)_d10_J01}
\end{figure}

The structure of the spectrum and the integrated density of states
of model (\ref{Hg}) have some features similar to those for the
isotropic model. The spectrum has the low- and high-energy regions
containing $6^{n}$ and ($8^{n}-6^{n}$) states. The gap $\Delta E$
between them decreases from the value $\Delta E=1+2g^{2}$ at $J=0$
to zero at $J\simeq 1$. Then $\Delta E$ increases and becomes
$\Delta E\simeq J$ at $J\gg 1$, i.e. the high-energy part of
spectrum goes to infinity at $J\to\infty $. The width $E_{w}$ of
the low-energy region is changed from $E_{w}\simeq J$ at small $J$
to $E_{w}\simeq 1$ at $J\gg 1$. The main difference between the
density of states for the cases $\Delta _{1}=1$ and $\Delta
_{1}\gg 1$ concerns the structure of the region with $6^{n}$
states. At $g=0$ this region corresponds to the low-energy part of
the Ising-Heisenberg model and it consists of $G_{0}(n)$ ground
states with zero energy and the part with energies $E\simeq J$ for
small $J$ and $E\simeq 1$ for $J\gg 1$. At $g>0$, $G_{0}(n)$
states are split as noted before into the states with $E=0$,
$E\sim g^{2}$, $E\sim g^{4}$ and so on as shown on
Fig.\ref{spectrum_Delta10} where the integrated density of states
of model (\ref{Hg}) with $n=6$ for $g=0.05$ and $J=0.01$ is
presented. As it can be seen from Fig.\ref{spectrum_Delta10} the
low-energy region (up to $E\sim g^{2}$) is separated by a gap from
the region with higher energy. The low-energy region below the gap
is divided into the parts with the energies from $E\sim g^{2}$ to
$E\sim g^{6}$. Such structure of the spectrum leads to the
specific features of the low-temperature thermodynamics. The
temperature dependence of the specific heat is shown in
Fig.\ref{C(T)_d10_J01}. The low-temperature maxima on the
dependence $C(T)$ are related to the corresponding parts of the
spectrum with the energies $E\sim g^{2}$, $E\sim g^{4}$ and $E\sim
g^{6}$. Two pronounced maxima at $T\simeq J$ and $T\simeq 1$ on
Fig.\ref{C(T)_d10_J01} are related to the parts of the spectrum
with $6^{6}$ and ($8^{6}-6^{6}$) states, correspondingly.
Similarly to the case $\Delta _{1}=1$ these maxima merge at
$J\simeq 1$ and are separated again at $J>1$. The high-temperature
maximum shifts to an infinity at $J\to\infty$. In the limit $J\to
0$ the region of the spectrum containing $6^{n}$ states has zero
energy and the specific heat has one maximum at $T\simeq 1$.

We note that the number of states in the low-energy region below
the gap is $G_{0}(6)=2042$. It means that this region consists of
both the ground states and the energy levels which are split off
from the ground state at $g=0$. We believe that this is a general
property of the strongly anisotropic model (\ref{Hg}) and the
number of the low-energy states below the gap is $G_{0}(n)$. We
note that this fact takes place for both spin-$\frac{1}{2}$
triangle chain and the delta-chain with spins $s_{1}=\frac{1}{2}$,
$s_{2}=1$.

However, such property (all state below the gap are split off from
the ground state at $g=0$ and the total number of these states is
$(3.56..)^{n}$ does not valid when $g$ increases. At first, the
gap in the spectrum is transformed to the quasi-gap or to the cusp
in the density of states as it is in the isotropic model
considered above. Secondly, the number of states below the cusp
for the isotropic model ($\Delta _{1}=1$) with $n=6$ is
$4^{6}=4096$ rather than $2042$. Therefore, the low-energy regions
determining the low-temperature thermodynamics are different for
the isotropic and the strong anisotropic spin-$\frac{1}{2}$
triangle chain and the number of states in the low-energy region
is $4^{n}$ for $\Delta _{1}=1$ and $(3.56..)^{n}$ for $\Delta
_{1}\gg 1$.

\section{Summary}

We have studied the ground state and the low-temperature
thermodynamics of the frustrated model consisting of the triangles
with competing ferro- and antiferromagnetic interactions. The
triangles are connected by the ferromagnetic Heisenberg
interaction $-J$. At definite conditions between values of the
interactions the lowest branch of the one-magnon excitations is
dispersionless (flat-band) which leads to the existence of exact
ground states which are isolated magnons and specifically
overlapping localized multi-magnon complexes. The ground state is
macroscopically degenerated in zero magnetic field and the
residual entropy is non-zero. Under these conditions the model
depends on one parameter and describes the transition line between
the ferromagnetic ground state and other ground state phases. The
ground state degeneracy does not depend on $J$ but the spectrum of
the excitations does. At $J=0$ the spectrum consists of isolated
bands separated by gaps. The lowest part of the spectrum contains
$6^{n}$ states and is separated from high-energy parts by the gap
depending on $J$. In turns, in this band there is the lowest
energy region of $4^{n}$ below the quasi-gap (a cusp in the
density of states). Therefore, it is possible to divide the
spectrum on three parts: high-energy part above the gap, the part
below the gap but above the quasi-gap and the low-energy region
below the quasi-gap. These three parts are responsible for the
peculiarities of the low-temperature thermodynamics of the
isotropic model. In particular, these three parts of the spectrum
are related to three maxima in $C(T)$ for small and large $J$. The
part containing $4^{n}$ states is related to weak peak in $C(T)$
and other two parts are responsible for broad maxima. At $J\simeq
1$ the gap between them vanishes and two pronounced maxima merge.
For $J\gg 1$ the gap is proportional to $J$ and the temperature of
the third maximum grows as $J$. As a result in the limit
$J\to\infty $ two maxima exist as for the F-AF delta chain to
which reduces the spin-$\frac{1}{2}$ triangle chain. At $J=0$
model consist of independent triangles and there is one maximum in
$C(T)$ at $T\simeq 1$.

In the limit of strong anisotropic interactions the spectrum has a
multi-scale structure which consists of subsets rank-ordered on
small parameter. Each subset is responsible for the appearance of
the peak in the temperature dependence of the specific heat.

We expect that similar properties of spin-$\frac{1}{2}$ triangle
chain can be present in other frustrated spin models, for example,
in linear chain of triangular F-AF plaquette with the
ferromagnetic Heisenberg interaction between them (kagome-like
chain).

\begin{acknowledgments}
The numerical calculations were carried out with use of the ALPS
libraries \cite{alps}.
\end{acknowledgments}

\end{document}